\documentclass[12pt]{article}
\input epsf
\newcommand{\be}{\begin{equation}}
\newcommand{\ee}{\end{equation}}
\newcommand{\bea}{\begin{eqnarray}}
\newcommand{\eea}{\end{eqnarray}}
\newcommand{\bean}{\begin{eqnarray*}}
\newcommand{\eean}{\end{eqnarray*}}
\font\upright=cmu10 scaled\magstep1
\font\sans=cmss12
\newcommand{\ssf}{\sans}
\newcommand{\stroke}{\vrule height8pt width0.4pt depth-0.1pt}
\newcommand{\Z}{\hbox{\upright\rlap{\ssf Z}\kern 2.7pt {\ssf Z}}}

\newcommand{\C}{{\rlap{\rlap{C}\kern 3.8pt\stroke}\phantom{C}}}
\newcommand{\R}{\hbox{\upright\rlap{I}\kern 1.7pt R}}
\newcommand{\CP}{\C{\upright\rlap{I}\kern 1.5pt P}}
\newcommand{\identity}{{\upright\rlap{1}\kern 2.0pt 1}}
\newcommand{\half}{\frac{1}{2}}

\newcommand{\z}{{\bar z}}
\newcommand{\y}{{\bar y}}

\begin{document}
\pagestyle{plain}

\title{\vskip -70pt
\begin{flushright}
{\normalsize DAMTP-2012-72} \\
\end{flushright}
\vskip 50pt
{\bf \Large \bf Vortex solutions of the Popov equations}
 \vskip 30pt
}
\author{Nicholas S. Manton\thanks{email: N.S.Manton@damtp.cam.ac.uk} \\[15pt]
{\normalsize
{\sl Department of Applied Mathematics and Theoretical Physics,}}\\
{\normalsize {\sl University of Cambridge,}}\\
{\normalsize {\sl Wilberforce Road, Cambridge CB3 0WA, England.}}\\
}
\vskip 20pt
\date{December 2012}
\maketitle
\vskip 20pt

\begin{abstract}
Popov recently discovered a modified version of the Bogomolny
equations for abelian Higgs vortices, and showed they were integrable
on a sphere of curvature $\half$. Here we construct a large family of
explicit solutions, where the vortex number is an even integer. There 
are also a few solutions without vortices. The solutions are 
constructed from rational functions on the sphere. 
\end{abstract}

\vskip 80pt

\newpage

In a recent paper \cite{Pop}, Popov discovered a novel set of
equations for vortices on a 2-sphere. They are a variant of the 
familiar Bogomolny equations for vortices in the U(1) abelian Higgs 
model \cite{Bog,JT,ManSut}. 

Recall that the usual abelian Higgs model on a surface
$\Sigma$, at critical coupling, can be obtained by dimensional 
reduction of the pure SU(2) Yang--Mills gauge theory on the
four-manifold $\Sigma \times S^2$. One imposes SO(3) symmetry on
the fields over $S^2$ and obtains a U(1) gauge theory over
$\Sigma$. The self-dual Yang--Mills equation on $\Sigma \times S^2$
reduces to Bogomolny vortex equations on $\Sigma$. Specially
interesting is where $\Sigma$ is the hyperbolic plane $H^2$, 
and the curvatures of $H^2$ and $S^2$ add to zero. The vortex 
equations simplify to Liouville's equation in this case, and
explicit solutions can be constructed using a class of 
holomorphic functions, as was first shown by Witten \cite{Wit}.

Popov turned this dimensional reduction around, starting with an
SU(1,1) Yang--Mills theory on the four-manifold $\Sigma \times
H^2$. Here one can impose SU(1,1) symmetry on the fields over
$H^2$. The self-dual Yang--Mills equation again reduces to vortex
equations over $\Sigma$, with gauge group U(1). It is somewhat accidental 
that the gauge group in four dimensions is the same as the symmetry group, 
but it needs to be non-compact for the dimensional reduction to work 
in a non-trivial way. Now the specially interesting, integrable case is where 
$\Sigma$ is $S^2$, with the curvatures of $S^2$ and $H^2$ again adding
to zero. So the four-manifold is the same as before, but the symmetry 
imposed on the fields is different, and the gauge group is different.

Let us fix the 2-sphere to have Gauss curvature $\half$, and hence
radius $\sqrt{2}$. Introducing a local complex coordinate $z$ in the
usual way, the metric is
\be
ds^2 = \frac{8}{(1 + z\z)^2} \, dz d\z \,,  
\ee
and the area form $\frac{8}{(1 + z\z)^2} \, \frac{i}{2} \, dz \wedge d\z$.
We will not discuss the four-geometry or the SU(1,1) Yang--Mills theory, 
and refer to Popov's paper \cite{Pop} for the details of this. We just
consider the vortex equations on $S^2$. These involve, locally, 
a complex scalar field $\phi$ and a U(1) gauge potential $a$, with 
complex components $a_z$ and $a_{\z} = {\overline{a_z}}$. Popov's 
equations (using the conventions of \cite{ManSut}) are
\bea
& & D_{\z} \phi \equiv \partial_{\z} \phi - i a_{\z} \phi = 0 \,, 
\label{Popov1} \\
& & F_{z \z} + \frac{2i}{(1 + z\z)^2}(1 - \phi \bar\phi) = 0 \,, 
\label{Popov2}
\eea
where $F_{z \z} = \partial_z a_{\z} - \partial_{\z} a_z$ and is
imaginary for a U(1) gauge potential\footnote{By exchanging $z$
and $\z$ one obtains equations for antivortices. These were analysed
in the first arXiv version of this paper.}.
The first equation is the same as for Bogomolny vortices on the
sphere, but for Bogomolny vortices, the second equation would be 
$F_{z \z} - \frac{2i}{(1 + z\z)^2}(1 - \phi \bar\phi) = 0$.

For this U(1) gauge theory on a 2-sphere, there is just one 
topological invariant, the first Chern number. This determines the 
topological class of the complex line bundle over the sphere, whose 
section is $\phi$ and whose connection form is $a$. The Chern 
number is the integral
\be
N = \frac{1}{2\pi}\int_{S^2} F_{z \z} \, dz \wedge d{\z} \,,
\label{Chern}
\ee
and is an integer. $N$ is also the net vortex number, which is defined 
for any smooth section $\phi$ with isolated zeros. Each zero is 
identified as a vortex whose multiplicity is the winding number of 
the phase of $\phi$ around a small circle (traversed once 
anticlockwise) enclosing the zero. The net vortex number is 
the sum of these multiplicities. The proof that it is equal to 
the first Chern number is purely topological, and uses the transition 
functions defining the U(1) bundle. It does not depend on 
any field equations. 

The Popov equations can be obtained from the energy function
\bea
E &=& \frac{1}{2}\int_{S^2}  \left\{ -\frac{(1 + z\z)^2}{2} F_{z \z} F_{z \z} 
-2(D_z\phi \overline{D_z\phi} + D_{\z}\phi \overline{D_{\z}\phi})\right. 
\nonumber \\
&& \qquad \qquad \qquad 
+ \left. \frac{2}{(1 + z\z)^2}(1 - \phi \bar\phi)^2 \right\} 
\frac{i}{2} \, dz \wedge d\z \,.
\eea
Because $F_{z \z}$ is imaginary, the first term is positive, as is the
last term. However, the middle terms are negative.
Making the usual Bogomolny rearrangement, dropping a total
derivative term, and using (\ref{Chern}), one finds
\bea
E &=& \half \int_{S^2} \Bigg\{ -\frac{(1 + z\z)^2}{2}
\left( F_{z \z} + \frac{2i}{(1 + z\z)^2}(1 - \phi \bar\phi)\right)^2
\nonumber \\
&& \qquad \qquad \qquad \qquad \qquad 
- \, 4 \, D_{\z}\phi \overline{D_{\z}\phi} \Bigg\} \frac{i}{2} 
\, dz \wedge d\z \ - \ \pi N \,.
\label{BogoEnergy}
\eea
Therefore, with $N$ fixed, the energy is stationary and has value 
$-\pi N$ for fields that satisfy the Popov equations (\ref{Popov1}) and 
(\ref{Popov2}). The Euler--Lagrange equations associated to $E$ are also
satisfied. However, the energy is not minimal, because the contributions
to the integral in (\ref{BogoEnergy}) are not all positive.

Now let us consider solutions of the Popov equations. One solution has $\phi$
identically zero. The connection $a$ is that of a Dirac monopole on $S^2$. 
Integrating eq.(\ref{Popov2}) one finds that $N=-2$, and although $N$ is
non-zero, there are no vortices. Another solution has $\phi =1$ and $a
= 0$. This is the trivial solution with $N=0$.

Using the $\bar\partial$-Poincar\'e lemma, Taubes proved that if 
$\phi$ is not identically zero and satisfies eq.(\ref{Popov1}), 
then its zeros, if it has any, are isolated and only vortices of positive 
multiplicity are possible. The reason is as follows. By complexifying 
the gauge group, one can go to the gauge where $a_{\z} = 0$ and 
then (\ref{Popov1}) says that $\phi$ is holomorphic. If $\phi$ has 
a zero at $z_0$, say, the leading term in its Taylor expansion 
is $A(z - z_0)^k$ for some positive integer $k$, so the 
multiplicity is $k$. It follows that $N$, the net vortex number and 
hence first Chern number, is positive or zero.
 
Integrating eq.(\ref{Popov2}) over the sphere gives the constraint
\be
\int_{S^2} \phi \bar\phi \, \frac{8}{(1 + z\z)^2} \, 
\frac{i}{2} \, dz \wedge d{\z} = 8\pi + 4\pi N \,,
\label{Bradlow}
\ee
where the first term on the right hand side is the area of the
sphere. The left hand side is non-negative, so $N \ge -2$. However, 
there is no constraint on the magnitude of $N$ if $N$ is positive. 
This constraint is very similar to that found by Bradlow for 
Bogomolny vortices \cite{Bra}. For Bogomolny vortices on a 2-sphere 
of area $8\pi$, the right hand side of (\ref{Bradlow}) is 
$8\pi - 4\pi N$, which restricts $N$ to be 1, with $N=0$ and $N=2$ 
also possible, but not giving true vortices. 

An immediate consequence of the above constraint is that for Popov
vortices, with positive $N$, $|\phi|$ cannot be limited to the 
range $0 \le |\phi| \le 1$ (as it is for Bogomolny vortices), because 
if it were, the integral on the left would be no greater than $8\pi$. 
This again distinguishes Popov and Bogomolny vortices.

To progress, we now eliminate the gauge potential to derive the 
analogue of the gauge invariant Taubes equation \cite{JT}. 
Equation (\ref{Popov1}) has the formal solution
\be
a_{\z} = -i\partial_{\z} \log\phi \,,
\label{a_zbar}
\ee
and therefore $a_z = i\partial_z{\log\bar\phi}$. 
It follows that $F_{z \z} = -i\partial_z \partial_{\z}
{\log(\phi\bar\phi)}$, so eq.(\ref{Popov2}) simplifies to
\be
\partial_z \partial_{\z}{\log(\phi\bar\phi)} =  
\frac{2}{(1 + z\z)^2}(1 - \phi \bar\phi) \,.
\ee
This is valid away from the zeros of $\phi$. Now writing $\phi
\bar\phi = |\phi|^2 = e^u$, we obtain
\be
4 \, \partial_z \partial_{\z} u = \frac{8}{(1 + z\z)^2}(1 - e^u) \,.
\label{TauPop}
\ee
This differs from the usual Taubes equation only by a change of sign
on the left hand side. The operator on the left is the flat Laplacian.
The equation implies that at a point where $u$ has a 
local maximum and a strictly negative Laplacian, $u$ must be 
positive. This is consistent with the need for regions where 
$|\phi| > 1$, and hence $u>0$. Strict local maxima of $|\phi|$ can 
only occur in these regions.

Equation (\ref{TauPop}) is a Liouville-type equation that can be solved 
explicitly. Rather than follow the argument of Witten, which 
demonstrated this for the usual Taubes equation on the hyperbolic 
plane (with the sign reversed on the left, and $(1 - z\z)^2$ in the 
metric factor), we follow the more geometric approach of ref. \cite{MR}. 

We start with the formula for the Gauss curvature $K$ of a general 
metric of the form $ds^2 = \Omega(z,\z) dz d{\z}$ on a Riemann surface,
\be
K = -\frac{2}{\Omega} \partial_z \partial_{\z}(\log \Omega) \,.
\ee
Next, we use this formula to calculate the Gauss curvature $K'$ of the
conformally related metric $ds^2 = e^{v(z,\z)} \Omega(z,\z) dz
d{\z}$. This is
\bea
K' &=&  -\frac{2}{e^v \Omega} \partial_z \partial_{\z}(v + \log \Omega) \\
   &=& \frac{1}{e^v}\left(-\frac{2}{\Omega} \partial_z \partial_{\z} v +
   K \right) \,.
\eea
Suppose now that both $K$ and $K'$ have the constant value
$\half$. Then 
\be
4 \, \partial_z \partial_{\z} v = \Omega(1 - e^v) \,.
\ee
If we set $\Omega = \frac{8}{(1 + z\z)^2}$, for which $K=\half$,
and if we set $v=u$, then this is just equation (\ref{TauPop}). The conclusion
is that if we can find a metric 
\be
ds^2 = e^{u(z,\z)} \frac{8}{(1 + z\z)^2} \, dz d{\z}
\label{umetric}
\ee 
with Gauss curvature $\half$, then $u$ is a solution of our vortex problem.

It is actually quite easy to find many metrics with this structure and 
with the desired curvature. One takes the metric on the 2-sphere once
more,
\be
ds^2 = \frac{8}{(1 + y\y)^2} \, dy d{\y} \,,
\ee
with complex coordinate $y$. This has Gauss curvature $\half$. Now
change coordinates by setting $y = R(z)$, which doesn't change the
curvature. The metric becomes
\be
ds^2 = \frac{8R'(z)\overline{R'(z)}}{(1 + R(z)\overline{R(z)})^2} \, 
dz d{\z} \,.
\ee
This is of the desired form (\ref{umetric}), with
\be
|\phi|^2 = e^u = 
\frac{R'(z)\overline{R'(z)}(1 + z\z)^2}{(1 + R(z)\overline{R(z)})^2} \,.
\label{phisquared}
\ee
$R(z)$ can be any meromorphic function, but to achieve a finite vortex
number and smooth fields over $S^2$, we must take $R(z)$ to be a rational 
function of $z$. The expression (\ref{phisquared}) solves eq.(\ref{TauPop}).
Notice that although $R'$ generally has poles, $|\phi|$ remains finite.

We have presented this construction in terms of a change of variable. More 
sophisticated is to say that the rational function $R$ is a holomorphic
map from the $z$-sphere to the $y$-sphere, given by the formula $y
= R(z)$. This is generally a ramified map (the inverse is a branched covering).
Suppose that $R(z)$ has degree $n$, i.e. is a ratio of polynomials in 
$z$ of degree $n$. Then the topological degree of the map from the sphere 
to itself is $n$, but this is not the vortex number. Vortices occur on 
the $z$-sphere where $|\phi|=0$, that is, at the ramification points 
where $R'(z) = 0$. Generically, there are $2n - 2$ such points, and 
they are simple zeros. The metric (\ref{umetric}) degenerates at these
points, and the curvature is singular, but this does not matter.

Given $|\phi|^2$, an appropriate local gauge choice for 
$\phi$ itself is
\be
\phi = \frac{R'(z)(1 + z\z)}{1 + R(z)\overline{R(z)}} \,.
\label{holophi}
\ee
The gauge potential $a_{\z}$, given by eq.(\ref{a_zbar}), is then
\be
a_{\z} = i\left(
\frac{R(z){\overline{R'(z)}}}{1 + R(z){\overline{R(z)}}} 
- \frac{z}{1 + z\z} \right) \,.
\ee
The zeros of $\phi$ are the zeros of $R'(z)$, and as $R'$ is
locally holomorphic, they have positive multiplicity. The vortex number 
is $N = 2n - 2$. The case $n=1$ is not excluded, but here $\phi$ has no zeros.

If the expression (\ref{holophi}) were globally smooth over $S^2$,
then the bundle would be trivial, in contradiction to the generally non-zero 
vortex number. In fact there are singularities, and of two
types. Assume that $R(z)$ is a generic rational function of degree $n$,
of the form
\be
R(z) = \frac{a_0 + a_1 z + \cdots + a_n z^n}{b_0 + b_1 z + \cdots + b_n z^n} 
\ee
with $a_0, a_n, b_0, b_n$ all non-zero, and the zeros and poles
all simple. The first type of singularity is a point $Z$ where
$R(Z)=\infty$. There are $n$ such points. Nearby, $R(z) \sim
c/(z-Z)$, so $\phi \sim c'(\z - {\bar Z)}/(z - Z)$. Although $|\phi|$
has a finite, non-zero value at $Z$, the phase of $\phi$ rotates by $-4\pi$
around $Z$. This phase rotation needs to be removed by a gauge
transformation on an annulus enclosing $Z$, of winding number 2. Summing 
over the $n$ points, the net winding number is $2n$. The second type 
occurs at $z = \infty$. In its neighbourhood, $R(z) \sim a + b/z$, so 
$\phi \sim b' \z/z$. Again, $|\phi|$ is non-zero,
and $\phi$ has some winding. The winding can be removed by a gauge 
transformation defined on an annulus on $S^2$ enclosing $z=\infty$, this time
of winding number $-2$. The conclusion is that $\phi$, as given by 
(\ref{holophi}), extends to a smooth section of a line bundle with 
Chern number $2n - 2$. This is consistent with the vortex number.

We have directly checked the constraint (\ref{Bradlow}) in the simple 
case that $R(z) = z^k$, with $k>1$. This rational function gives a vortex 
solution on $S^2$ with circular symmetry, and reflection symmetry in the 
equator $|z| = 1$. There are vortices of multiplicity $k-1$ at $z=0$ and
$z=\infty$. The total vortex number is $N = 2k - 2$. The integral on
the left hand side of (\ref{Bradlow}) is elementary, and equal to
$8\pi k$ as expected from the right hand side. $|\phi|$ has its 
maximum value on the equator $|z|=1$, where $|\phi| = k$. This 
value is greater than 1, as we argued earlier it had to be. 

Note that there are also non-trivial solutions with $N=0$. For
example, if $R(z) = cz$, with $c$ real and positive, then
\be
\phi = \frac{c(1 + z\z)}{1 + c^2z\z} \,.
\ee
$\phi$ is real, circularly symmetric, and has no zeros and no
winding. If $c>1$, then $\phi$ decreases monotonically from $c$ at $z=0$ 
to $1/c$ at $z = \infty$. For $c=1$, this is the trivial solution 
with $\phi = 1$ everywhere and $F_{z\z} = 0$.

Finally, let us look at the moduli space of these solutions. The space
of rational functions of degree $n$ has real dimension $4n + 2$. However,
an SU(2) M\"obius transformation, which is a degree 1 rational map, 
is an isometry of the metric on $S^2$. Composing a rational map
$y = R(z)$ with such a M\"obius transformation on the $y$-sphere has
no effect on the fields. Therefore, our construction leads to a moduli
space of vortices of dimension $4n - 1$. Since the vortex number is $N
= 2n - 2$, the dimension can be re-expressed as $2N + 3$. 

In conclusion, in the usual abelian Higgs model, the Bogomolny equations are
integrable when the underlying surface is the hyperbolic plane with
curvature $-\half$. Popov's abelian vortex equations, which differ 
only slightly, are integrable on a 2-sphere with curvature
$\half$. Here we have shown how to construct explicit vortex solutions 
using rational functions on the sphere, and have given them 
a geometric interpretation in terms of conformal rescalings of the 
2-sphere metric that preserve the curvature except at the vortex 
locations. For the solution obtained from a rational function 
$R(z)$ of degree $n$, the vortex locations are the 
points where $R'(z)=0$. There are $2n-2$ such points, all of positive
multiplicity, so the vortex number is $N=2n-2$, generally an even, 
positive integer. There are also trivial and non-trivial solutions 
with $N=0$ and no vortices, and a solution with $N=-2$ and $\phi$ 
vanishing identically. It would be interesting to determine if solutions 
with odd vortex numbers are possible.

\section*{Acknowledgement}

I am grateful to Steffen Krusch for a discussion about these vortices.

\end{document}